\documentclass[showpacs,amsmath,amssymb,onecolumn]{revtex4}

\usepackage{graphicx}
\usepackage{dcolumn}

\begin{document}

\title{The optical phonon spectrum of SmFeAsO}
\author{C. Marini$^{1,2}$, C. Mirri$^{1}$, G. Profeta$^{3}$, S. Lupi$^{1}$, D. Di Castro$^{1}$,
R. Sopracase$^{1}$, P. Postorino$^{1}$, P. Calvani$^{1}$, A.
Perucchi$^{4}$, S. Massidda$^{5}$, G. M. Tropeano$^{6}$, M.
Putti$^{6}$, A. Martinelli$^{7}$, A. Palenzona$^{7}$ and P.
Dore$^{1}$}
\affiliation{$^1$ CNR-INFM  ``Coherentia" and Dipartimento di Fisica, Universit\'{a} di Roma ``La Sapienza", P.le A.Moro 2, 00185 Rome, Italy}
\affiliation{$^2$ ``Unit\'{a} CNISM Roma1" - Dipartimento di Fisica ``E. Amaldi", Universit\'{a} degli Studi Roma Tre, via della Vasca Navale 84,00146 Rome, Italy}
\affiliation{$^3$ CNISM - Dipartimento di Fisica, Universit\'{a} degli Studi dell'Aquila, Via Vetoio 10, I-67010 Coppito (L'Aquila), Italy}
\affiliation{$^4$ Sincrotrone Trieste S.C.p.A., S.S. 14 km 163.5, in Area Science Park, 34012 Basovizza (Trieste), Italy}
\affiliation{$^5$ Universit\'{a} degli Studi di Cagliari, S.P. Monserrato-Sestu km 0.700, I-09124 Monserrato (Cagliari), Italy}
\affiliation{$^6$ CNR-INFM-LAMIA and Dipartimento di Fisica, Via Dodecaneso 33, 16146 Genova, Italy}
\affiliation{$^7$ CNR-INFM-LAMIA and Dipartimento di Chimica e Chimica Industriale,  Via Dodecaneso 31, 16146 Genova, Italy}

\begin{abstract}
We measured the Raman and the Infrared phonon spectrum of
SmFeAsO polycrystalline samples. We also performed Density
Functional Theory calculations within the pseudopotential
approximation to obtain the structural and dynamical lattice
properties of both the SmFeAsO and the prototype LaFeAsO compounds.
The measured Raman and Infrared phonon frequencies are well
predicted by the optical phonon frequencies computed at the $\Gamma$
point, showing the capability of the employed \emph{ab-initio}
methods to describe the dynamical properties of these materials. A
comparison among the phonon frequencies of different oxypnictides
suggests a possible role of the high frequency phonons in the
pairing mechanism leading to superconductivity in these materials.
\end{abstract}

\date{\today}

\maketitle

\section{Introduction}
The recent discovery of high temperature superconductivity in
iron-based oxypnictides \cite{kamih} has triggered in the scientific
community a strong research effort on this new class of
superconductors. Superconducting pnictides are layered compounds
containing a transition metal (Fe, Co, Ni) bonded with a group V
element (As and P) and intercalated with layers formed by a
rare-earth (\emph{RE} = La, Sm, Nd, Ce) and oxygen.
Superconductivity appears upon doping with electrons (substitution
of O with F) or with holes (oxygen deficiency), while the parent
compounds show an antiferromagnetic metallic ground state. After the
discovery of superconductivity at 26 K in F-doped LaFeAsO
\cite{kamih}, higher critical temperatures were soon obtained by
substituting La with Sm \cite{xhchen}, Ce \cite{gfchen}, Pr
\cite{ren4283} and Nd \cite{renEurLet}. Also oxygen-free compounds
like \textit{A}Fe$_2$As$_2$ (\textit{A} = Ca, Sr, Ba \cite{rotter,sasmal}) were found to
exhibit superconductivity, even if at lower temperatures, thus
opening the way to a new 'iron-age' \cite{grant}.

The prototype LaFeAsO shows an anomaly near $T_{SDW}$ = 150 K both
in the resistivity and in the magnetic susceptibility, which is
attributed to a spin-density-wave (SDW) instability \cite{dong,
yild, klauss}. Just above $T_{SDW}$, the system undergoes a
structural transition from a high temperature tetragonal (P4/nmm)
phase to an orthorhombic (Cmma) phase \cite{delacruz, nomura}. It is
generally recognized that doped compounds become superconducting
when electron or hole doping suppresses the SDW
instability\cite{huang}. However, a satisfactory theory of the
superconducting mechanism in the pnictides has not yet been found.
In particular, the role of phonons in the pairing mechanism has been
strongly questioned due to the small value of the electron-phonon
coupling\cite{boeri}. The scenario is also complicated by
difficulties in an \emph{ab-initio} description of the normal state
properties of these compounds \cite{mazin}. A careful comparison
between experimental and calculated properties is thus timely and
important in the validation of the present first-principle
theoretical methods. Structural and dynamical properties of the
\emph{RE}FeAsO compounds are an obvious ground where this comparison
is due.

In this context we measured both the Raman and the Infrared (IR)
phonon spectrum of pure SmFeAsO polycrystalline samples, well
characterized in terms of crystal structure (P4/nmm at high
temperature as in SmFeAsO), magnetic, thermal and transport
properties showing that $T_{SDW} \sim$ 135$\div$140 K
\cite{martin1,tropeano}. We have then calculated the phonon spectrum
and compared the theoretical results with those of the experiments.

\section{Raman and Infrared spectra}

Raman spectra were measured in back-scattering geometry using a
confocal micro-Raman spectrometer equipped with a He-Ne laser at
632.8 nm. Measurements were performed on fresh-cut surfaces without
any additional treatment. A camera coupled to the microscope allowed
to select the point over the sample surface on which to focus the
laser spot (about 5 $\mu$m by using a 20X objective). Since SmFeAsO
is a very poor Raman scatterer \cite{had}, spectra with a good
signal-to-noise ratio were collected within the 80-1000 cm$^{-1}$
range by setting long acquisition times and a large number of
spectra to be averaged. It was verified that the obtained results
are not affected by laser-induced sample heating. The
polycrystalline nature of the sample prevent any polarization
analysis of the collected spectra.

\begin{figure}[h]
\includegraphics[width=12cm]{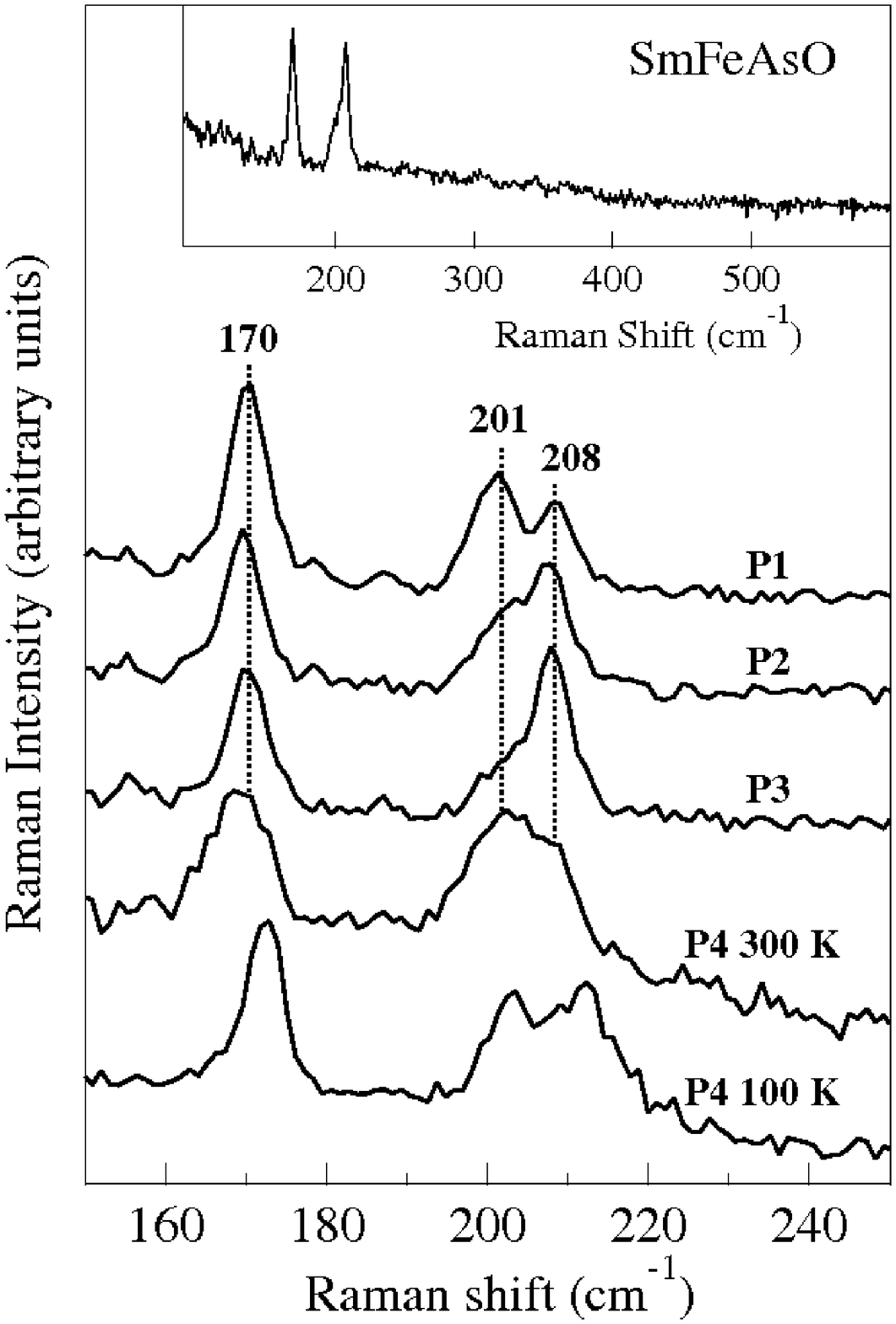}\caption{Raman spectra collected
between 150 and 250 cm$^{-1}$ in different sample points (P1-P4) at 300 K.
The 100 K spectrum collected at point P4 is shown for comparison.
Vertical dashed lines mark the room temperature peak frequency
values. In the inset, a typical room temperature Raman spectrum of
SmFeAsO up to 600 cm$^{-1}$ is shown.} \label{fig.1}
\end{figure}

A typical room temperature Raman spectrum  is shown in the inset of
Fig \ref{fig.1}, where Raman peaks are evident between 170 and 210
cm$^{-1}$. Only in a few cases the collected spectra show Raman
contributions due to samarium oxide \cite{hongo}, in agreement with
the results of the analyses performed on the same sample
\cite{martin1}. Spectra collected from different points (P1-P4) are
reported in Fig. \ref{fig.1} in a limited frequency range to
highlight the details of the Raman peaks. Three lines are evident,
at frequencies (170, 201 and 208 cm$^{-1}$) which coincide with
those reported in a previous room temperature Raman study which
includes polarization analysis \cite{had}. The different relative
intensities of the peaks we observed in different points must be
ascribed to polarization effects. Indeed the intensity of the Raman
peaks can depend on the angle between the laser polarization and the
possible preferred orientation of the crystallites in the scattering
volume.

We also studied the temperature dependence of the Raman spectrum
between 300 and 100 K by mounting the sample in a cryostat, which
requires longer acquisition times. The comparison between the
spectra collected from the same point (P4) at 300 and 100 K (see
Fig. \ref{fig.1}) shows no remarkable changes in the number of
phonons or in the peak frequencies occurs decreasing temperature below T$_{SDW}$,
besides the expected phonon hardening and narrowing. The 300 and 100
K peak frequency values are reported in Table \ref{table1}.

The infrared reflectivity R($\omega$) of the same SmFeAsO sample was
then measured from 50 to 5000 cm$^{-1}$ and between 17 and 300 K.
The reference was the sample itself, gold-coated by evaporation
in-situ. These measurements, performed using a Michelson
interferometer at the SISSI Infrared beamline of the ELETTRA Storage
Ring (Trieste, Italy), showed that R($\omega$) is independent of
temperature above 1000 cm$^{-1}$. The reflectivity measurements were
then extended at up to 30000 cm$^{-1}$ at room temperature, by using
a monochromator and an aluminum film as reference.

\begin{figure}[h]
\includegraphics[width=12 cm]{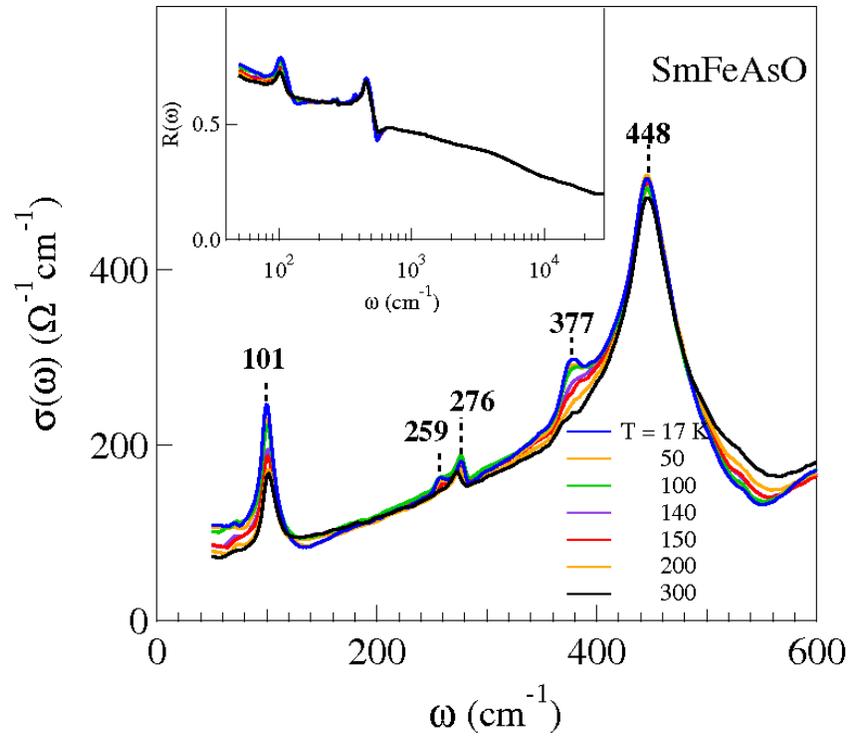}\caption{(Color on line) Optical
conductivity at selected temperatures in the phonon region. The 100
K peak frequency values are reported. The inset shows the
reflectivity R($\omega$) of SmFeAsO in the whole frequency range at
the same temperatures} \label{fig.2}
\end{figure}

The inset of Fig. \ref{fig.2} shows that, as expected, R($\omega$)
has a bad metallic character. A Hagen-Rubens relation was then
employed for the extrapolation to $\omega$ = 0, usual power laws for
that to $\omega$ = $\infty$, and the optical conductivity
$\sigma_1(\omega)$ was then obtained by standard Kramers-Kronig
transformations. Due to the wide frequency range where R($\omega$)
is measured directly, $\sigma_1(\omega)$ in the phonon frequency range
(see Fig. \ref{fig.2}) does not depend appreciably on the choice of
the extrapolations.

Although the present paper is devoted to the study of the optical
phonons, it is worth noting that, in agreement with previous results
on LaFeAsO \cite{dong}, the optical conductivity of SmFeAsO does not
provide a clear evidence of a SDW gap, indicating that not all of the Fermi
surface is gapped. Only an increase of the low frequency
conductivity owing to the increase of carrier mobility is observed
for decreasing temperature below $T_{SDW}$ \cite{puttilupi}.

In the far-infrared spectrum five phonon peaks at 101, 259, 276,
377, and 448 cm$^{-1}$ are evident at low temperatures. Above 150 K,
the 259 and 375 cm$^{-1}$ peaks are hardly discernible and a
standard Lorentzian fit was employed to determine their peak
frequencies. As in the Raman case, no remarkable change in the peak
frequencies occurs on decreasing temperature. The 300 and 100 K peak
frequencies are reported in Table \ref{table1}. We note that the
structural phase transition around $T_{SDW}$ from tetragonal to
orthorhombic recently observed in SmFeAsO\cite{martin2}, as well as
in LaFeAsO\cite{delacruz, nomura} and in NdFeAsO\cite{bianconi},
does not significantly affect the phonon frequencies, as also
observed in the LaFeAsO case\cite{dong, yild}. However, the
significant increase in the relative intensities of the 259 and 377
cm$^{-1}$ peaks below 150 K may be regarded as an effect of the
structural transition around $T_{SDW}$.

\section{Calculations}
The experimental data reported in Table \ref{table1} were compared
with Density Functional Theory (DFT) calculations of the SmFeAsO
optical phonons, performed within the pseudopotential approximation
for the electron-ion interaction\cite{bloch, kresfurt, kresjou}. The
f-states of Sm were considered to be inside the core and a formal
valence \emph{+3} was assumed for Sm (\emph{$6s^25d^1$} as valence
electrons, with \emph{5s} and \emph{5p} treated as semi-core
states). The electronic density, integrated on a uniform mesh of
$14\times 14\times 10$ k-points, converged within an energy cutoff
of 400 eV. A local-density approximation with gradient
corrections\cite{perdew} was used for the exchange correlation
energy.

For comparison and cross checking with previous linear-response
theory calculations, we also performed calculations for the
prototype LaFeAsO compound, in the same P4/nmm structure as SmFeAsO.
The theoretical lattice constants were obtained by minimizing the
total energy as a function of the volume, after relaxing all the
internal degrees of freedom. In all calculations, both systems are
considered in the paramagnetic phase.

The resulting lattice constants for LaFeAsO are a = 4.017 \AA, c =
8.677 \AA, $z_{La}$ = 0.144 and $z_{As}$ = 0.638, in good agreement
with previous computations\cite{boeri}, while for SmFeAsO we find a
= 3.948 \AA, c = 8.336 \AA, $z_{Sm}$ = 0.141 and $z_{As}$ = 0.647,
in reasonable agreement with the experimental findings\cite{martin1}
(for a discussion on the agreement between experimental and computed
parameters see Ref \cite{mazin2,Leb}). We notice that the smaller Sm
ion reduces both the lattice constants: the in-plane one is reduced
by about 2\%, while the out-of-plane one is reduced by 4\%. This
reduction does not reflect on all interatomic bond lengths. The most
affected parameter is the height of \emph{RE} atoms above the O
plane, passing from 1.25 \AA$ $ in LaFeAsO to 1.17 \AA$ $ in
SmFeAsO. Consequently, the \emph{RE}-O bond length is reduced from
2.36 \AA$ $  to 2.30 \AA. The Fe-As bond lenghts instead changes
from 2.34 \AA$ $ in LaFeAsO to 2.36 \AA$ $ in SmFeAsO, although the
As height above the Fe plane increases upon Sm substitution. As we
will see, these structural properties are crucial to explain the
differences in the phonon frequencies between these compounds.

The phonon frequencies at $\Gamma$, to be compared with the measured
frequencies, were obtained by diagonalizing the dynamical matrix
calculated in the frozen phonon approach\cite{alfe}. For each phonon
mode we report in Table \ref{table1} experimental and computed frequency, the
irreducible representation associated with the mode, the
polarization (X,Y: in-plane, Z: out-of-plane), the optical activity
(IR or Raman), and the dominant atoms involved. A comparison with
the present calculations allows one to assign all the observed
phonon lines. We also report the calculated frequencies in LaFeAsO,
in good agreement with previous computations \cite{yild, boeri}, and
the corresponding experimental values from literature data
\cite{dong, had}.

A close inspection of data in Table \ref{table1} shows a good
agreement (very good in some cases) between the measured and
calculated frequencies. As expected, the low-frequency part of the
spectrum (up to about 210 cm$^{-1}$) is dominated by vibrations
mainly involving the massive atoms (\emph{RE} and As). The
comparison with the calculated frequencies in LaFeAsO reveals a
small but systematic increase of the frequencies associated with
in-plane motions. The larger Sm mass with respect to La would
account for a slight reduction of the frequencies
($\omega_{Sm}/\omega_{La} \simeq (M_{La}/M_{Sm})^\frac{1}{2}$) in
passing from La to Sm, but this is washed out by the reduction of
the strong 2D in-plane bonds. The same happens for the mode with
Z-polarization (97 cm$^{-1}$) that also involves the Fe-As motion.
On the other hand, the modes at 177 and 203 cm$^{-1}$ (out-of-plane
vibrations of \emph{RE} and As) turn out to be softer than the
respective ones in LaFeAsO, as partially confirmed by the
experiments.

It is worth noting that, due to the weakness of the \emph{RE}-As
bond, the ionic interactions between the layers are weaker than the
in-layer ones.We then expect that a simple rescaling of the
frequencies with the \emph{RE} mass should be valid. This was
verified by using the force constants calculated in the case of
LaFeAsO (at its own lattice constants) and diagonalizing the
dynamical matrix with the La atomic mass substituting that of Sm. We
found that the softening observed in SmFeAsO is exactly predicted by
this model calculation for only these two modes.

At frequencies up to 300 cm$^{-1}$ the observed modes only involve
the Fe-As network. The mode at 224 cm$^{-1}$ involves only the Fe
atoms moving along the Z-direction, and thus is expected to be
unaffected by the Sm substitution and by the slight Fe-As bond
reduction. The other modes, Fe-As stretching modes, are consistently
higher in passing from LaFeAsO to SmFeAsO.

Only in the region above 300 cm$^{-1}$, which is dominated by the
vibrations involving the oxygen atoms, the phonon frequencies of the
two compounds are significantly different. In fact, the substitution
of La with Sm produces a shortening of the \emph{RE}-O bond length
and of the $z_{\emph{RE}}$ parameter (see the discussion above).
This produces an overall stiffening of the \emph{RE}-O modes; in
particular, the highest frequency E$_g$ phonon is at 434 cm$^{-1}$
in LaFeAsO, at 503 cm$^{-1}$ in SmFeAsO.

\section{Conclusion}
We performed Raman and IR measurement on SmFeAsO, thus obtaining
accurate information on the optical phonon frequencies at the
$\Gamma$ point. They show that the structural phase transition
around 140 K does not significantly affect the phonon frequencies,
as previously observed in the LaFeAsO case. We also performed a
theoretical study of the structural and dynamical lattice properties
by means of accurate Density Functional Theory calculations within
the pseudopotential approximation. Our calculations show that, in
both SmFeAsO and LaFeAsO, the Raman and IR phonon frequencies are
well predicted by pseudo-potential density functional methods. This
confirms the capability of \emph{ab-initio} methods to describe the
dynamical properties of these materials.

We find that the phonon frequencies in the two materials are very
similar at low $\omega$, while those at high $\omega$ are much
higher in SmFeAsO than in LaFeAsO. In particular, the E$_g$ oxygen
mode is predicted to be at 434 cm$^{-1}$ in LaFeAsO, at 503 cm$^{-1}$  in
SmFeAsO. We have shown that this is related to the \emph{RE} atom
being closer to the oxygen plane in SmFeAsO than in LaFeAsO. As a
general property, the same effect can be anticipated in other
oxypnictides. In fact, results on other \emph{RE}FeAsO compounds
reported in the literature show that the same mode is observed at
450 cm$^-1$ for \emph{RE}=Ce\cite{zhao}, at 485 cm$^-1$ for
\emph{RE}=Nd\cite{zhang}. Bearing in mind that the
superconducting transition temperature T$_c$ of the related F-doped
systems at optimum doping depends on the \emph{RE} element (T$_c$ =
25 K for \emph{RE} = La\cite{kamih}, T$_c$ = 40 K for \emph{RE} =
Ce\cite{zhao,ren4283}, T$_c$ = 51 K for \emph{RE} =
Nd\cite{renEurLet}, and T$_c$ = 55 K for \emph{RE} =
Sm\cite{xhchen}) and that the phonon frequencies are not strongly
affected by a low level F-doping\cite{zhang}, the increase of T$_c$
appears to be correlated to the increase of the E$_g$ phonon
frequency. Although the electron-phonon coupling constant is very
low for LaFeAsO\cite{boeri}, and probably for all oxypicnitides, the
above comparison suggests that a role of the phonons in the pairing
mechanism cannot be simply ruled out, especially in view of the
complex and strong magneto-phonon interaction \cite{yin} expected in
these materials.

\section{Acknowledgements}
Work partially supported by the Italian Ministry of Education,
through PRIN 200602174 project, by the Italian MIUR Project Cybersar
(Consorzio COSMOLAB, PON 2000-2006), and  by INFM-CNR through a
supercomputing grant at Cineca (Bologna, Italy). One of authors
(P.D.) would like to thank L. Boeri for fruitful discussions.

\newpage 

\begin{table}
\label{table1} \vspace{0.5cm} 
\begin{tabular}{cccccccc}
SmFeAsO &   SmFeAsO &   LaFeAsO &   LaFeAsO &       &       &       &   Involved atoms  \\
Exp.    &   Calculated  &   Calculated  &   Exp. \cite{had, dong}    &   Simmetry    &   Polarization    &   Activity    &   (\emph{RE} = Sm , La)   \\
\hline
    &   68  &   64  &       &   Eu  &   X,Y &   IR  &   \emph{RE}+0+Fe+As  \\
    &       &       &       &       &       &       &       \\
101 &   97  &   89  &   97  &   A2u &   Z   &   IR  &   \emph{RE} +Fe+As   \\
(101)    &       &       &       &       &       &       &       \\
    &   118 &   117 &       &   Eg  &   X,Y &   RAM &   \emph{RE} +As  \\
    &       &       &       &       &       &       &       \\
    &   148 &   144 &       &   Eg  &   X,Y &   RAM &   \emph{RE} +Fe+As   \\
    &       &       &       &       &       &       &       \\
170 &   177 &   184 &   162 &   A1g &   Z   &   RAM &   \emph{RE} +As  \\
(172)    &       &       &       &       &       &       &       \\
201 &   203 &   205 &   208 &   A1g &   Z   &   RAM &   \emph{RE} +As  \\
(203)    &       &       &       &       &       &       &       \\
208 &   224 &   224 &   201 &   B1g &   Z   &   RAM &   Fe  \\
(212)    &       &       &       &       &       &       &       \\
259 &   269 &   257 &   248 &   A2u &   Z   &   IR  &   Fe+As   \\
(259)    &       &       &       &       &       &       &       \\
273 &   283 &   276 &   266 &   Eu  &   X,Y &   IR  &   Fe+As   \\
(276)    &       &       &       &       &       &       &       \\
    &   300 &   287 &       &   Eg  &   X,Y &   RAM &   Fe+As   \\
    &       &       &       &       &       &       &       \\
345*    &   350 &   291 &   316 &   B1g &   Z   &   RAM &   O   \\
    &       &       &       &       &       &       &       \\
377 &   361 &   298 &   336 &   Eu  &   X,Y &   IR  &   \emph{RE} +O   \\
(377)    &       &       &       &       &       &       &       \\
446 &   435 &   412 &   430 &   A2u &   Z   &   IR  &   \emph{RE}+O+As \\
(448)    &       &       &       &       &       &       &       \\
    &   503 &   434 &       &   Eg  &   X,Y &   RAM &   O   \\
    &       &       &       &       &       &       &       \\
\end{tabular}
\caption{Room temperature peak frequencies in cm$^{-1}$ for the
Raman and infrared phonons of SmFeAsO, Present room temperature
Raman data coincide with those of Ref. \cite{had}, where also the
Raman line at 345 cm$^{-1}$ (marked by *) is reported. Peak
frequencies at 100 K are given in brackets. Calculated phonon
frequencies are given for SmFeAsO and LaFeAsO, the latter compared
with literature experimental values from Refs. \cite{had, dong}.
Mode symmetry, polarization (X,X: in-plane, Z: out-of-plane),
optical activity (IR or Raman) and dominant atoms involved in each
mode are also given.}

\end{table}
\normalsize


\begin{thebibliography}{27}
\bibitem{kamih} Y. Kamihara, T. Watanabe, M. Hirano and H. Hosono, J. Am. Chem. Soc.\textbf{130}, 3296(2008).

\bibitem{xhchen}X. H. Chen, T. Wu, G. Wu, R. H. Liu,  H. Chen and D. F. Fang, Nature \textbf{453}, 761(2008).

\bibitem{gfchen} G. F. Chen, Z. Li,  D. Wu, G. Li, W. Z. Hu, J. Dong, P. Zheng,  J. L. Luo and  N. L. Wang, Phys. Rev. Lett. \textbf{100},24700 (2008).

\bibitem{ren4283} Z. A. Ren, J. Yang, W. Lu,  W. Yi, G. C. Che,  X. L. Dong, L. L. Sun and Z.-X. Zhao, Materials Research Innovations \textbf{12}, 105 (2008).

\bibitem{renEurLet} Z. A. Ren, J. Yang, W. Lu, W. Yi, X. L. Shen, Z. C. Li, G. C. Che, X. L. Dong, L. L. Sun, F. Zhou and  Z. X. Zhao, Europhys. Lett. \textbf{82},57002 (2008).

\bibitem{rotter} M. Rotter, M. Tegel and  D. Johrendt, Phys. Rev. Lett. \textbf{101}, 107006 (2008).

\bibitem{sasmal}  K. Sasmal, B. Lv,  B. Lorenz,  A. M. Guloy, F. Chen,  Y. Y. Xue and C. W. Chu, Phys. Rev. Lett. \textbf{101}, 107007 (2008).

\bibitem{grant} P. Grant, Nature \textbf{453}, 1000 (2008).

\bibitem{dong}  J. Dong, H. J. Zhang,  G. Xu, Z. Li, Li G.,  W. Z. Hu, D. Wu, G.F. Chen, X. Dai, J. L. Luo, Z. Fang and N. L. Wang, Europhys. Lett. \textbf{83}, 27006 (2008).

\bibitem{yild} T. Yildirim T., Phys. Rev. Lett. \textbf{101}, 057010 (2008).

\bibitem{klauss} H. H. Klauss, H. Luetkens, R. Klingeler, C. Hess,  F. J. Litterst,  M. Kraken,  M. Korshunov, I. Eremin,  S. L. Drechsler, R. Khasanov, A. Amato,  J. Hamann-Borrero, N. Leps,  A. Kondrat, G. Behr, J. Werner and B. Bchner, Phys. Rev. Lett. \textbf{101}, 077005 (2008).

\bibitem{delacruz}  C. De la Cruz, Q. Huang,  J. W. Lynn, J. Li, W. Ratcliff, J. L. Zarestky, H. A. Mook, G. Chen, J. L. Luo, N. L. Wang and  P. Dai, Nature 453, 899 (2008).

\bibitem{nomura}T. Nomura, S. W. Kim, Y. Kamihara, M. Hirano, P.V. Sushko, K. Kato, M. Takata, A. L. Shluger and H. Hosono, arXiv:0804.3569(2008).

\bibitem{huang} Q. Huang, J. Zhao, J. W. Lynn, G. F. Chen, J. L. Luo, N.
L. Wang and P. Dai, Phys. Rev. B \textbf{78},  054529 (2008).

\bibitem{boeri} L.Boeri, O.V. Dolgov and A. A. Golubov, Phys. Rev. Lett. \textbf{101}, 026403(2008).

\bibitem{mazin} I. I. Mazin, M. D. Johannes, L. Boeri, K. Koepernik and D. J. Singh, Phys. Rev. B \textbf{78}, 085104 (2008).


\bibitem{martin1}
A. Martinelli, M. Ferretti, P. Manfrinetti, A. Palenzona, M. Tropeano, M. R. Cimberle, C. Ferdeghini, C. Valle, C. Bernini, M. Putti and A. S. Siri, Supercond. Sci. Technol.
\textbf{21}, 095017 (2008).

\bibitem{tropeano} M. Tropeano, A. Martinelli, A. Palenzona, E. Bellingeri, E. Galleani dAgliano, T. D. Nguyen, M. Affronte and M. Putti, Phys. Rev. B \textbf{78}, 094518 (2008).


\bibitem{had} V. G. Hadjiev, M. N. Iliev, K. Sasmal, Y. Y. Sun and C. W. Chu, arXiv:0804.2285(2008).

\bibitem{hongo} T. Hongo,  K. Kondo, K. Nakamura and T. Atou, J. Mater. Sci. \textbf{42}, 2582 (2007).

\bibitem{puttilupi} M. Tropeano,  C. Fanciulli, C. Ferdeghini, D. Marr, A. S. Siri,  M. Putti, A. Martinelli, M. Ferretti,  A. Palenzona, M. R. Cimberle,  C. Mirri, S. Lupi,  R. Sopracase,  P. Calvani and  A. Perucchi, arXiv:0809.3500 (2008).
\bibitem{martin2}
  A. Martinelli, A. Palenzona, C. Ferdeghini, M. Putti  and E. Emerich arXiv:0808.1024(2008), in press on Journal of Alloys and
Compounds.

\bibitem{bianconi} M. Fratini, R. Caivano, A. Puri,A. Ricci, Z. Ren, X. Dong, J. Yang, W. Lu, Z. Zhao, L. Barba, G. Arrighetti, M. Polentarutti and A. Bianconi,
Supercond. Sci. Technol. \textbf{21}, 092002 (2008).

\bibitem{bloch} P. Bl\"ochl, Phys. Rev. B \textbf{50}, 17953 (1994).

\bibitem{kresfurt} G. Kresse and J. Furthm\"uller, Phys. Rev. B \textbf{54}, 11169 (1996).

\bibitem{kresjou} G. Kresse  and D. Joubert, Phys. Rev. B \textbf{59}, 1758 (1999).

\bibitem{perdew} J. Perdew and  Y. Wang, Phys. Rev. B \textbf{45}, 13244 (1992).

\bibitem{mazin2} I.I. Mazin,  M. D. Johannes, L. Boeri, K. Koepernik and D.J.Singh, arXiv0806.1869 (2008).

\bibitem{Leb} S. Leb\"{e}gue, P. Yin and  W. E. Pickett, arXiv0810.0376 (2008).


\bibitem{alfe} D. Alf\'{e}, G. Price, and M. Gillan, Phys. Rev. B \textbf{64}, 04512316 (2001).

\bibitem{zhao} S. C. Zhao,  D. Hou, Y. Wu,  T. L. Xia,  A. M. Zhang,  G. F. Chen, J. L. Luo,  N. L. Wang,  J. H. Wei,  Z. Y. Lu and  Q. M. Zhang, arXiv0806.0885 (2008)

\bibitem{zhang} L.Zhang, T. Fujita,  F. Chen,  D. L. Feng,  S. Maekawa and  M. W. Chen, arXiv0809.1474 (2008)

\bibitem{yin} Z. Yin, S. Leb\"{e}gue, M. Han, B. Neal, S. Savrasov, and W. Pickett, Phys. Rev. Lett. \textbf{101}, 047001 (2008).

\end{thebibliography}
\end{document}